# Searching for the optimum conditions for silicene growth by calculations of the free energy

Yu-Peng Liu[a, b], Bo-Yuan Ning[c], Le-Cheng Gong[a, b], Tsu-Chien Weng[c] and Xi-Jing Ning[a, b]*

[a]*Institute of Modern Physics, Fudan University, Shanghai, China;* [b]*Applied Ion Beam Physics Laboratory, Fudan University, Shanghai, China;* [c]*Center for High Pressure Science & Technology Advanced Research, Shanghai, China*

Xi-Jing Ning, Institute of Modern Physics, Applied Ion Beam Physics Laboratory, Fudan University, Shanghai 200433, People's Republic of China

E-mail: xjning@fudan.edu.cn

# Searching for the optimum conditions for silicene growth by calculations of the free energy

Very recently we developed an efficient method to calculate the free energy of 2D materials on substrates and achieved high calculation precision for graphene or γ-graphyne on copper substrates. In the present work, the method was further confirmed to be accurate by molecular dynamic simulations of silicene on Ag substrate using empirical potential and was applied to predict the optimum conditions based on *ab initio* calculations for silicene growth on Ag (110) and Ag (111) surface, which are in good agreement with previous experimental observations.

Impact Statement: This work paves a way based on *ab initio* calculations to predict the optimum conditions for various 2D materials growth by deposition of the atoms on substrates.

## 1. Introduction

Silicene, as a 2D analogy of graphene, has recently aroused widespread concerns [1-10]. However, the preparation is more difficult than graphene since there exist no corresponding layered bulk material in nature like graphite to be exfoliated for obtaining the sheet. Recently, vapour deposition (VD) method was applied to grow silicene on metal substrate, showing that the surface structure significantly affects the growth. As an example on silver substrate, the growth on the (001) surface just produced a "complex" superstructure without clear symmetry[11] and on the (110) surface, silicon nanoribbons (NRs) formed along the [$\bar{1}$10] direction with honeycomb structure [12,13], while on the (111) surface, a continuous graphenelike 2D honeycomb arrangement of silicon atom, silicene, can be obtained [1,14]. It is notable that the silicene can grow just when temperature of the substrate was kept between 220 and 250℃, showing that the temperature has quite significant effect on the growth.

Clearly, lots of experimental efforts could be saved for preparing a desired 2D materials by VD, such as silicene, if the suitable surface of a substrate and the optimum temperature could be predicted theoretically. The predictions might be made in principle by calculations of the free energy (FE), but the calculation on condensed matter has been an open problem since the born of statistical physics by the end of 19th century, although some progress has taken place in the past 30 years[15-21]. Very recently, we developed a direct integral approach (DIA) for calculating the FE of 2D material on substrate, and the high calculation precision was confirmed by molecular dynamic (MD) simulations of graphene and γ-graphyne on Cu (111) surface at temperature up to 1300 K[22]. The method works at least 3-orders faster than state-of-the-art algorithm and enables us to obtain the free energy via ab initio calculations.

In the present work, the DIA was applied to calculate the FE (or partition function, PF) of silicene on Ag substrate at temperature up to 1300K. Firstly, we employed empirical potentials (Tersoff and Morse potential) to describe the atomistic interaction to test the accuracy of DIA on this system by comparing the internal energy derived from the PF with the MD simulations. Then ab initio calculations based on density function theory (DFT) were performed to describe the atomistic interaction for calculating the PF (or FE) by the DIA. Based on the calculated FE of a piece of silicene with different morphology on the (110) or (111) surface of Ag substrate at temperature up to 2000K, the optimum conditions (specific surface at certain temperature) for the growth of NRs or sheets were predicted, which are in good agreement with the experimental observations.

## 2. Accuracy of DIA

The model for a 2D materials of N atoms on a substrate is shown in Fig.1, where the substrate of M atoms is treated as a thermal bath at temperature T. The total potential is expressed as

$$U(x^{3N}, X^{3M}) = U_{2D}(x^{3N}) + V(x^{3N}, X^{3M}), \qquad (1)$$

where $U_{2D}$ is the potential energy of the 2D material with the coordinates of the atoms denoted by $x^{3N}(x_1, x_2 ... x_{3N})$, and V is the interaction potential between the 2D material and the substrate with its atoms denoted by $X^{3M}(X_1, X_2 ... X_{3M})$. The PF of 2D materials can be expressed as,

$$Z = \frac{1}{N!} \left(\frac{2\pi m}{\beta h^2}\right)^{\frac{3N}{2}} Q, \qquad (2)$$

where $\beta = 1/k_B T$ with kB the Boltzmann factor, and Q is the configurational integral

$$Q = \int d x^{3N} \exp[-\beta U(x^{3N}, X^{3M})]. \qquad (3)$$

By introducing a new set of coordinates, $x'_i = x_i - q_i$, where $q_i$ is coordinates of the atoms in state of the lowest potential energy $U_0$, Eq. (3) can be rewritten as

$$Q = e^{-\beta U_0} \int d x'^{3N} \exp[-U'(x'^{3N}, X^{3M})], \qquad (4)$$

where $U'(x'^{3N}, X^{3M})$ is defined by

$$U'(x'^{3N}, X^{3M}) = U(x^{3N}, X^{3M}) - U_0. \qquad (5)$$

According to our previous work [22],

$$Q = e^{-\beta U_0} \prod_{i=1}^{3N} L_i, \qquad (6)$$

where the effective length $L_i$ is defined as

$$L_i = \int e^{-\beta U'(0...x'_i...0, X^{3M})} dx'_i. \qquad (7)$$

To obtain the effective lengths, the first step is to find the most stable structure of the 2D materials with lowest potential $U_0$, which can be accomplished by common algorithms of geometrical optimization or dynamic damping methods[23,24]. Starting from the most stable structure, atom $i$ of the 2D material is moved step by step in one degree of the freedoms, such as $x'_i$, while $y'_i$ and $z'_i$ as well as all other atoms keep fixed to determine $U'(0 ... x'_i ... 0, X^{3M})$.

For a piece of silicene sheet of 336 Si atoms on the (111) surface of an Ag substrate with 2640 Ag atoms arranged in perfect fcc lattices (Fig. 1(a)), Tersoff potential[25] was taken to represent the interactions between Si atoms, and the interaction between the Si atoms and the Ag atoms is described by Morse pairwise function[26]. The system was cooled down to 0.01K by a damping method[23] to determine the lowest energy $U_0$ and the most stable structure. Although $U'(0 ... x'_i ... 0, X^{3M})$ felt by an atom i located at the edges of the silicene sheet should be different from the one felt by an atom in the center region, the edged atoms are much fewer than the center atoms, so we just only calculated the $U'$ felt by a center atom (Fig. 1 (b)) to determine the effective length $L_x$, $L_y$ and $L_z$, and the configuration integral approximates to

$$Q = e^{-\beta U_0}[L_x L_y L_z]^N. \tag{8}$$

Applying Eq. (2) and $E = -\frac{\partial}{\partial \beta} \ln Z$, the internal energy ($E_{PF}$) can be obtained through

$$E_{PF} = \frac{3}{2}Nk_B T + \frac{k_B T^2}{Q}\frac{\Delta Q}{\Delta T} \tag{9}$$

with $\Delta T = 0.1 K$.

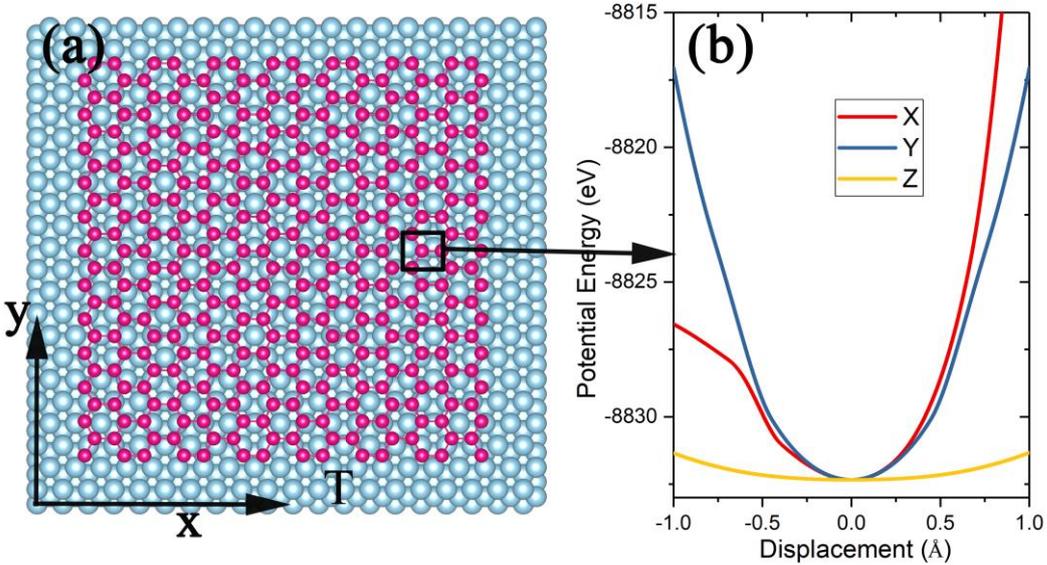

**Figure 1.** Schematic of a piece of silicene of N atoms (red balls) lying on the surface of Ag(111) substrate of M atoms (blue balls) at temperature T (a), and the potential energy (b) felt by a Si atom (show in black box) moving along the X-, Y- or Z-axis.

In order to test the accuracy of the above algorithm, a classical MD simulations was performed to produce the internal energy of the system with the same potentials [25,26] as the ones for calculations of the PF. All the simulations were implemented in a Large-scale Atomic/Molecular Massively Parallel Simulator (LAMMPS) program. Specifically, the time step is set as 0.1 *fs*, and velocities of the silicon atoms are assigned according to a time integration on Nose-Hoover style non-Hamiltonian equations of motion to keep the system at a given temperature. Then the internal energy ($E_{MD}$) and the temperature were recorded every 30 *fs* to do the average over 100 records.

As shown in Fig. 2, the internal energy ($E_{PF}$) derived from the PF is in excellent agreement with that ($E_{MD}$) obtained by the MD simulations. For the temperatures lower than 500K, the relative error ($\frac{|E_{PF}-E_{MD}|}{|E_{MD}|} \times 100\%$) is only 0.001% and gradually increases up to 0.021% for 1300 K. It may be expected that the relative error would get smaller if the effective length $L_i$ of the edged atoms was calculated, instead of replacing $L_i$ with the one of the center atom, for obtaining Q by Eq. (6). It should be pointed out that such a test is much more stringent than the comparisons between the results derived from PF and experiments because the same interaction potential is used in both calculation of the PF and the MD simulations while the potential may not correctly describe the realistic interactions between atoms concerned in the experiment.

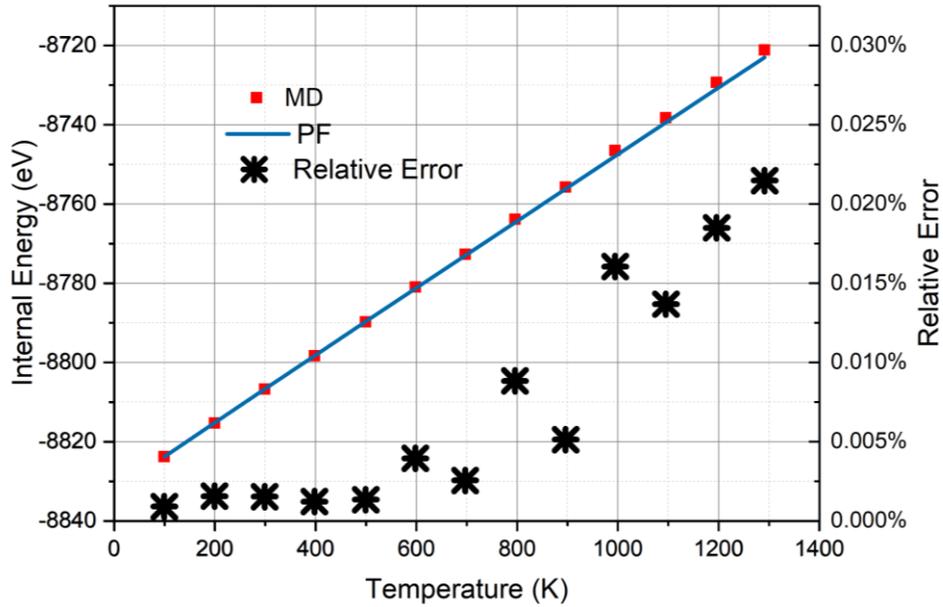

**Figure 2.** The internal energy as the function of temperature derived from the PF (blue line) or MD simulation (red square), where the relative errors $\frac{|E_{PF}-E_{MD}|}{|E_{MD}|} \times 100\%$ (black stars) are characterized at the right vertical axis.

## 3. Growth of silicene on Ag (110) and Ag (111) surface

In order to search for the optimum growth conditions for a given silicene grain on Ag substrate, we calculated the FE ($F = -k_B T \ln Z$) by DIA with the potential $U'$ determined by density functional theory (DFT), which were performed by Vienna *ab initio* Simulation Package (VASP) based on local density approximation (LDA). The kinetic-energy cutoff for the plane-wave basis set is 400 eV, and the Brillouin zone was sampled with $(2 \times 2 \times 1)$ *k*-points.

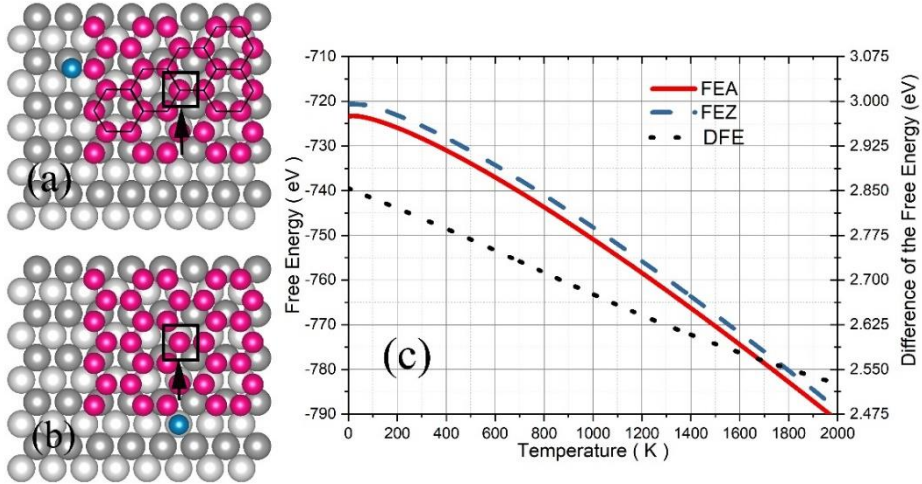

**Figure 3.** A silicene grain of four silicon hexagons on the Ag (110) surface with a deposited Si atom adhered to the zigzag (a) or armchair edge (b). The free energy for the zigzag adherence and armchair adherence and the difference are shown in (c).

Considering if silicene can grow on Ag (110) surface by deposition Si atom, we calculate the FE of a silicene grain consisting of 35 Si atoms in hexagon arrangement with a Si atom adhered to the zigzag (Fig. 3a) or armchair edge (Fig. 3b) on the surface of four atomic layer with each containing 40 Ag arranged in a $8 \times 5$ fcc supercell. According to thermodynamics, if the free energy for the zigzag adherence (FEZ) equates to the one for the armchair adherence (FEA), then the grain can develop both the zigzag and armchair edge to grow larger. Otherwise, the future Si atoms deposited on the Ag surface would more favorably arrive at the zigzag (or armchair) edge if FEZ is lower (or larger) than FEA, resulting in development of only one of the edges to form a band (or NR).

For calculations of FEZ and FEA, the system was optimized firstly with the bottom layer of the Ag substrate fixed, and then the adhered Si atom (and one of the Si atom denoted by a square in Fig. 3(a) and (b) in the center of the silicene) was moved along the X, Y and Z direction step by step with an interval of 0.1Å to produce the $U'$ for calculating the effective length $l_x$, $l_y$ and $l_z$ (or the $L_x$, $L_y$ and $L_z$) via Eq. (7), and finally, the configuration integral was obtained by

$$Q = e^{-\beta U_0}(l_x l_y l_z)(L_x L_y L_z)^{N_0}, \qquad (10)$$

where $N_0$ =35, the number of the Si atoms except for the adhered Si atom.

As shown in Fig. 3 (c), FEZ and FEA decrease significantly with the temperature up to 2000 K, while the difference DFE (= FEZ-FEA) decreases gradually from 2.854 eV for 0 K down to 2.526 eV for 2000 K. According to thermodynamics, if a Si atom deposited on this surface has enough time to wander between the two edges, then it will finally locate at the armchair edge because of the lower FE, and five similar Si atoms form an armchair edge instead of a zigzag one. As the results, the armchair edge progresses row by row and the initial silicene grain develop into a nanoribbon along the [$\bar{1}$11] direction with a width about 1.6 nm, which is right the observation in a previous experiment [12,13].

For the growth on Ag (111) surface, a silicene grain of 25 Si atoms arranged in hexagons with a Si atom adhered at the zigzag edge (Fig. 4(a)) or the armchair edge (Fig. 4(b)) was placed on a Ag substrate of four (111) layers with each of 48 Ag atoms in a (6×8) supercell, and the geometry was optimized with the bottom layer of the Ag substrate fixed. The adhered Si atom (and a Si atom denoted by a square in Fig. 4(a), (b)) was step by step to produce the potential $U'$, and the configuration integral was obtained by Eq. (10) with $N_0$=25. As shown in Fig. 4(c), the FEZ and FEA decrease with the temperature up to 2000 K, while the DFE increases slightly from -0.022 eV up to 0.056 eV. In the temperature range from 200 K to 580K the DFE is less than 0.01 eV, and the DFE equates to zero for 400 K. According to thermal dynamics, 400 K is the most optimum temperature for the silicene grain growing larger on both the zigzag and armchair edge to form a continuous graphene like structure. In the previous experiment [1,14], the optimum temperature is 493~580 K, which is about one hundred higher than our prediction. The difference may result from the limited calculations precision of DFT on such system.

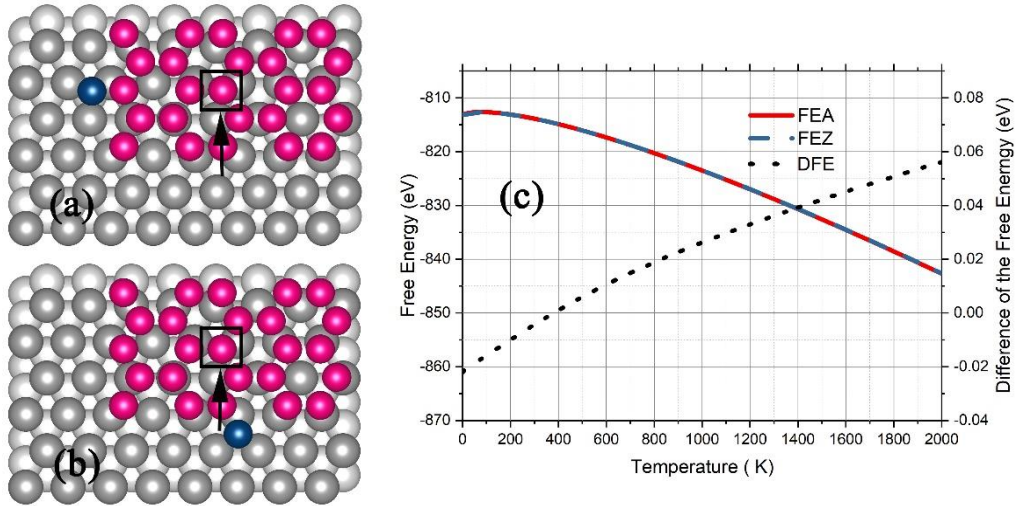

**Figure 4.** A silicene grain of four silicon hexagons on the Ag (111) surface with a deposited Si atom adhered to the zigzag (a) or armchair edge (b). The free energy for the zigzag adherence and armchair adherence and the difference are shown in (c).

## 4. Summary

In summary, an approach for calculating the free energy (or partition function) of 2D material on substrate was further confirmed by classical MD simulations of silicene on Ag substrate, and was applied to search for the optimum conditions for silicene growth by *ab initio* calculations on silver substrate. Based on the calculated free energy for several silicene grains on Ag surface it was shown that the (111) surface is suitable for silicene growth by deposition of Si atoms while the (110) can only grow nanoribbons, which are in good agreement with previous experimental observations. This work paves a way to predict optimum conditions for preparing various 2D material by deposition of the atoms on substrates.

**Disclosure statement**

No potential conflict of interest was reported by the authors.

**Funding**

The work is supported by National Natural Science Foundation of China under Grant No.21727801 and No.11274073.